\documentclass{aa}
\usepackage{graphicx}
\usepackage{txfonts}
\usepackage[figuresright]{rotating}

\begin{document}
   \title{High-resolution UVES/VLT spectra of White Dwarfs observed
   for the ESO SN Ia Progenitor Survey. II. DB and DBA
   Stars\thanks{Based on data obtained at the Paranal Observatory of
   the European Southern Observatory for programmes 165.H-0588 and
   167.D-0407.}}

\titlerunning{High-Resolution Spectra Observed for SPY. II. DB and DBA Stars}

   \author{B. Voss
          \inst{1}
          \and
	  D. Koester
          \inst{2}
	  \and
	  R. Napiwotzki
	  \inst{3}
	  \and
	  N. Christlieb
	  \inst{4}
	  \and
	  D. Reimers
	  \inst{4}
          }


   \institute{Astronomisches Rechen-Institut, Zentrum f\"ur Astronomie
der Universit\"at Heidelberg, M\"onchhofstr. 12-14, 69120 Heidelberg,
Germany\\ \email{voss@ari.uni-heidelberg.de} 
\and Institut f\"ur Theoretische Physik und Astrophysik, Universit\"at
Kiel, Leibnizstra\ss e 15, 24098 Kiel, Germany 
\and Centre for Astrophysics Research, University of Hertfordshire,
College Lane, Hatfield AL10 9AB, UK 
\and Hamburger Sternwarte, Gojenbergsweg 112, 21029 Hamburg, Germany }

   \date{Received Feb 30, 2007; accepted May 30, 2007}

   \abstract{We present a detailed spectroscopic analysis of the stars
   with helium-dominated spectra in the ESO Supernova Ia Progenitor
   Survey (SPY).}
   {Atmospheric parameters, masses, and abundances of
   trace hydrogen are determined and discussed in the context of
   spectral evolution of white dwarfs.}
   {The spectra are compared with theoretical model atmospheres using a
   $\chi^2$ fitting technique, leading to determinations of effective
   temperature, surface gravity, and hydrogen abundance.}
   {Our final sample contains 71 objects, of which 6 are new
   detections and 14 are reclassified from DB to DBA because of the
   presence of H lines. One is a cool DO with weak \ion{He}{ii} lines, 2 are
   composite DB+dM. 55\% of the DB sample show hydrogen and are thus
   DBA, a significantly higher fraction than found before.}
   {The large incidence of DBA, and the derived total hydrogen masses
   are compatible with the scenario that DBs ``reappear'' around
   30\,000\,K from the DB gap by mixing and diluting a thin hydrogen
   layer of the order of $10^{-15}\,M_{\odot}$. This hydrogen mass is then
   during the evolution continuously increased by interstellar
   accretion. There are indications that the accretion rate increases
   smoothly with age or decreasing temperature, a trend which
   continuous even below the current low temperature limit (Dufour
   2006). 
   A remaining mystery is the low accretion rate of H
   compared to that of Ca observed in the DBZA, but a stellar wind
   extending down to the lowest temperatures with decreasing strength
   might be part of the solution.}


   \maketitle

\section{Introduction}

About 20\% of all white dwarfs (WD) have atmospheres in which helium
is the main constituent. Such hydrogen-deficient stars are believed to
form when a cooling stellar remnant experiences a very late
helium-shell flash and goes through a born-again episode, i.e., a last
thermal pulse, in which almost all residual hydrogen is burnt (see,
e.g., Althaus et al. \cite{althaus}, and references therein).

As these stars advance on the WD cooling sequence, they can initially
be observed at temperatures down to approximately 45\,000\,K as
hot DO stars with spectra dominated by \ion{He}{ii} lines, and
subsequently as DB stars whose spectra show only \ion{He}{i}, at
temperatures that are mostly below 30\,000\,K. Finally, when Helium
becomes spectroscopically invisible below effective temperatures
around 10\,000\,K, the helium atmospheres display either a featureless
DC-type spectrum, or a spectrum dominated by the features of trace
constituents, i.e., dredged-up carbon leading to DQ spectra or
accreted metals leading to a DZ-type spectrum. Besides these main
types, DB spectra which show traces of hydrogen or metals are known,
classified as DBA or DBZ stars.
 
While the properties of the hydrogen-atmosphere DA white dwarfs as a
group have been closely investigated in numerous studies, the DBs
received less attention. They are less numerous than the DAs and thus
it is more difficult to assemble data sets that allow meaningful
statistical analyses. The most comprehensive study of DB stars so far
was that of a sample of 80 DBs, carried out by Beauchamp et
al. (\cite{beauchamp96}).

Since Beauchamp's study, the number of known DB stars was not 
significantly increased until two surveys, the Sloan Digital Sky
Survey (SDSS, York et al. \cite{york}) and the Supernova Ia Progenitor
Survey (SPY, Napiwotzki et al. \cite{napi03}) collected new samples of
DB spectra. The SPY sample contains high-resolution spectra of about
60 DB white dwarfs. These data allow to determine well-constrained
parameters of a large number of DB stars for first time since
Beauchamp's work. The SDSS spectra, on the other hand, are of medium
resolution, but much more numerous. Thus, the SDSS detected a large
number of new DBs. The latest version of the catalog of WDs in the
SDSS (Eisenstein et al. \cite{eisenstein_a}), based on the SDSS data
release 4 (DR4), contains 713 DB stars. About 560 of these stars are
ordinary DBs which only exhibit \ion{He}{i} in their spectra. The
second largest subgroup is that of DBA stars with weak Balmer lines in
the otherwise \ion{He}{i}-dominated spectrum; about 90 such stars are
found in the SDSS sample. Many of these newly detected DBs are faint,
which makes detailed spectroscopic studies difficult. Nevertheless,
the almost tenfold increase of the number of DB stars has allowed the
first statistically meaningful analysis of some key questions
connected to the evolution of DB stars:
 
As a helium-atmosphere WD cools, it encounters two temperature regions
of special interest, the ``DB gap'' between temperatures of 45\,000\,K
and 30\,000\,K, and later the instability region of the V777\,Her
variables near temperatures of 25\,000\,K. The DB gap region has been
enigmatic for some time since no DB stars were found in this range of
temperatures, raising questions about the very nature of
helium-atmosphere white dwarfs. However before the SDSS, only few cool
DO and hot DB stars that constrained the gap were known and thus the
reality of the gap was not well established. Eisenstein et
al. (\cite{eisenstein_b}) found several hot DB stars from the SDSS
within the DB gap, some of which exhibit weak \ion{He}{ii} lines and
are thus situated in the higher-temperature part of the DB gap. The
gap has however not been closed by their discovery because they also
find that the number density of DB stars in the gap is still lower by
a factor of about 2.5 than what would be expected from the density of
cooler DB stars.

As a possible explanation, it has been speculated that cooling DA
stars might change into DBs, provided that their hydrogen layer is
very thin, on the order of $M_{\mathrm{H}}\sim 10^{-15}M_{\odot}$. In
this case, the superficial hydrogen would be convectively mixed into
the more massive Helium layer when the atmosphere cools to
temperatures lower than 30\,000\,K and the Helium layer becomes
convective (see Eisenstein et al. \cite{eisenstein_b}, and references
therein). Some peculiar objects of mixed helium/hydrogen atmosphere
compositions, the DAB and some of the DBA stars, might be objects
which are currently undergoing such a transition and could therefore
provide insight into the viability of this scenario.

The majority of cool DBA stars however can more easily be explained as
DB stars that are accreting hydrogen from the ISM. They were long
found to comprise about 20\% of the DB stars, a fraction that was
first estimated by Shipman, Liebert \& Green (\cite{shipman}) and
later confirmed, e.g, by Beauchamp et
al. (\cite{beauchamp96}). Similarly, 22\% of all DB stars in the
catalogue of McCook \& Sion (\cite{WD}) are classified as DBA, and in
the recent SDSS DR4 sample, the fraction is about 13\%. It is clear
that the fraction of DBA among the DB stars is a question of the
observational hydrogen detection limits.

It has long been expected by some authors that in fact all DBs accrete
small traces of hydrogen and would thus appear as DBA if only small
enough traces of hydrogen could be detected. Most studies so far,
including the SDSS, could only detect hydrogen abundances, relative to
helium, larger than a few times $10^{-5}$. These abundance
measurements were mostly based on the appearance of H$\gamma$ and
H$\beta$ in the spectra. Lower abundances can be identified from the
Ly${\alpha}$ line in UV data (e.g. Provencal et al. \cite{provencal}),
but this method is more involved and not easily applied to large
samples. Hunter et al. (\cite{Hunter}) demonstrated that also optical
spectra allow a measurement of lower hydrogen abundances if they cover
the H$\alpha$ line. In their sample of 24 DB objects, the DBA fraction
is raised to about 30\%. With observations of a better $S/N$ and
higher spectral resolution, this fraction was expected to be even
larger, with the expectation that probably hydrogen might be
identified in a majority of the DB stars.

Fontaine et al. (\cite{Fontaine05}) proposed that stellar winds of WD
stars might prevent hydrogen accretion at temperatures in excess of
20\,000\,K, thus enabling the conversion from DB to DBA stars only at
lower temperatures. Observations with increased sensitivity that allow
to detect increasingly lower H masses in DB stars should allow to test
this effect and to provide new insight into the evolution of DBA
stars.

The high resolution WD spectra that were obtained by SPY contain the
largest sample to date of high-quality DB spectra and it should thus
be possible to determine lower hydrogen abundances than was possible
before from optical data.

The SPY data will also be helpful to investigate the question if
invisibly small amounts of trace hydrogen are present in hot DB stars
with temperatures above 20\,000\,K. This question is of importance
because even such small traces of hydrogen significantly affect the
shape of the \ion{He}{i} lines and thus complicate the determination
of temperatures from the spectra. Beauchamp et
al. (\cite{beauchamp99}) compared temperature estimates based on pure
He atmospheres to estimates including effects of invisibly small
hydrogen contaminations, and found a shift of up to 3000\,K in
temperature. The temperatures of hot DBs are therefore not well
constrained so far. An improvement of these temperature estimations
would greatly benefit the analysis of the V777\,Her instability strip.

\section{The SPY data} 

SPY is a radial velocity survey that was conducted to test the double
degenerate channel of the formation of supernovae Ia. About 800 white
dwarfs were observed in the course of the survey, assembling a large
collection of high-quality white dwarf spectra. Most of the DB stars
among the SPY objects were selected as targets based on the WD catalog
(McCook \& Sion \cite{WD}) and from a study of DBs in the Hamburg ESO
Survey by Christlieb et al. (\cite{christlieb}). A more detailed
discussion of the target selection for the SPY observations is given
in Napiwotzki et al. (\cite{Napi2001}, \cite{Napi2007}).

The SPY spectra were employed for a number of studies beyond the
original scope of SPY, of which the work of Koester et
al. (\cite{paper1}), hereafter ``paper I'', is most relevant to our
analysis. They have derived temperatures and gravities from a
preliminary sample of about 200 SPY stars, including 13 DB stars.

Since they have already given a detailed description of the data
properties, we only briefly summarize the observations and data
reduction:

The spectra were obtained with UVES, a high resolution echelle
spectrograph at the ESO VLT telescope. UVES was used in a dichroic
mode, resulting in small gaps, $\approx80$\,\AA~wide, at 4580\,\AA~and
5640\,\AA~in the final merged spectrum. The spectral resolution at
H$\alpha$ is $R=18\,500$ or better, and the $S/N$ per binned pixel
(0.05\,\AA) is $S/N=15$ or higher.

The spectra were reduced with the ESO pipeline for UVES, including the
merging of the echelle orders and the wavelength calibration. Paper I
found that the quality of these automatically extracted spectra is
very good, except for a quasi-periodic wave-like pattern that occurs
in some of the spectra. This has since been largely removed by
additional processing by collaborators of the SPY project at the
University of Erlangen-N\"urnberg. A detailed description of the
reduction procedure applied to the spectra analyzed here will be
published in Napiwotzki et al. (\cite{Napi2007}). Some artifacts
remain in the data, but they do not significantly affect the spectral
analysis.

For our analysis, the data have been rebinned to a step size of
0.1\,\AA. This is in contrast to the analysis of paper I which used a
binning of 1\,\AA. Therefore, we include the paper I subsample in our
analysis, now with the smaller step size. Further details about the
observational setup and the data reduction can be found in Napiwotzki
et al. (\cite{napi03}) and in paper I.

\section{Analysis method}

\subsection{Model atmosphere fits}
We fit the data with model spectra from six different grids of helium
atmosphere models with different hydrogen abundances, from H/He
$=10^{-2}$ down to H/He $=10^{-7}$, and with a grid of pure He
atmosphere models. The grids cover a range of temperatures from
10\,000\,K to 50\,000\,K and logarithms of the surface gravities from
7.0 to 9.0. The input physics of the models are described in Finley et
al. (\cite{FKB}), Homeier et al. (\cite{homeier98}), Koester \& Wolff
(\cite{koesterwolff}), and Koester et al. (\cite{koester2005a}). We
use a $\chi^2$ minimization fitting routine that is based on the
Levenberg-Marquardt algorithm (Press et al. \cite{press}) to
derive the best fitting effective temperature and surface gravity for
each spectrum. Some more details on the fitting process can be found
in Homeier et al. (\cite{homeier98}).

The SPY spectra are not reliably flux calibrated, and therefore we do
not use their continuum slopes for the fitting but only the profiles
of the helium lines. We determine a reference continuum level at each
wavelength to which the models are scaled; this reference level is
derived by interpolating the flux levels measured next to the
\ion{He}{i} lines.

The Helium lines reach their maximum strength around
$T_{\mathrm{eff}}=20\,000\,$K, and therefore two solutions for the fit
of a spectrum are normally possible, one on either side of the line
strength maximum. We employed two $T_{\mathrm {eff}}$ starting values
for the fitting, 15\,000\,K and 25\,000\,K, to cover all possible
solutions. 

For most objects it is not difficult to pick the correct
solution; if the temperature difference of both solutions is large,
the line profiles are different enough to allow an easy decision for
the best fit. Furthermore, for many objects there is only one
solution, either because both fits converge to a single solution, or
because the fit from the hot starting value converges to exceedingly
high values of the temperature. For the remaining few objects, a
careful study of the fit quality was necessary to determine which
solution is to be preferred; for some objects the hydrogen abundance
or the presence of pulsations provided additional constraints.

There are two stars in our sample that show composite spectra. The red
part of the spectra was not used in the fits of these DB+dM binaries;
only the \ion{He}{i} lines at short wavelengths of
$\lambda<4500$\,\AA\ were fitted, where the flux of the late-type
companion is negligible.

Below effective temperatures of approximately 15\,500 to 16\,500\,K --
depending on surface gravity -- neutral He van der Waals broadening
becomes more and more important for the line shapes. Since our
original model grids included only Stark broadening,
we have calculated a second grid up to 18\,000\,K which
includes Stark broadening in a simplified form and van der Waals
broadening combined. This greatly improved the agreement with the
observed line profiles and strengths. Because the fit was not perfect,
and also because in general the dependence of line profiles on surface
gravity is smaller for the cooler DBs, we have decided to keep $\log
g$ fixed at 8.0, in order to get a more robust estimate of the
effective temperature. A fit in which $\log g$ is allowed to vary
would converge to very high gravity values.

There are six objects in our sample whose spectra only show a strongly
broadened 5876\,\AA~line, or only this line together with about
equally strong H$\alpha$. These are very cool objects with helium-rich
atmospheres; they are related to peculiar stars like, e.g.,
HS\,0146+1847 (Koester et al. \cite{koester2005a}), which are
dominated by helium but appear hydrogen-rich because of their very low
temperature. Since only one or two lines are available in these
spectra, no meaningful fit would be possible and thus we estimate the
temperature and the hydrogen abundance from the equivalent widths of
\ion{He}{i} 5876\,\AA, and, if present, H$\alpha$, assuming a fixed
$\log g=8.0$.

\subsection{Determination of H/He}
We obtain seven different fit results for each observed spectrum, one
fit result per assumed hydrogen abundance. If a visual inspection of
the observed spectrum does not reveal the presence of H$\alpha$, we
adopt the pure helium atmosphere solution.

Otherwise, if H$\alpha$ is unambiguously present, as shown in
Figs. \ref{AllDBA} through \ref{AllDBA3}, we measure the equivalent
width of the line in the observed spectrum as well as in each of the
six model spectra. Then, we interpolate the model equivalent widths
$EW$, model temperatures $T_{\mathrm{eff}}$, and model gravities $\log
g$ as functions of the assumed hydrogen abundance H/He: $EW$(H/He),
$T_{\mathrm{eff}}$(H/He), and $\log g$(H/He). With the observed
equivalent width $EW_{\mathrm{obs}}$, we can determine
H/He$_{\mathrm{obs}}$ from the inversion of $EW$(H/He). We then adopt
$T_{\mathrm{eff}}($H/He$_{\mathrm{obs}})$ and $\log
g($H/He$_{\mathrm{obs}})$ as the temperature and gravity of the
object.

For most stars, the H$\alpha$ equivalent width does not rise
monotonously over the whole range of modeled abundances. Instead, it
mostly rises monotonously up to H/He\,$=10^{-3}$ and decreases again
for H/He\,$=10^{-2}$. The reason for this effect is that a fit of the
H/He\,$=10^{-2}$ models to the data results in a high temperature, for
which the equivalent width is small in spite of the high abundance. A
given equivalent width thus often allows two solutions for the H
abundance. For all but three of our objects the low-abundance solution
proves to be the better fit of the data. For the objects for which
instead the high abundance solution is preferred, the fit quality
difference of both solutions is small. Therefore it has to be noted
that for these objects a second solution of almost identical quality
exists.

\subsection{Stellar masses and hydrogen mass fractions}

The masses were determined from $T_{\mathrm{eff}}$ and $\log g$, and
the evolutionary mass-radius relations of Wood (\cite{wood}) for
``thin'' layers, appropriate for DB white dwarfs.

The depth of the convection zones in DB white dwarfs for given mass
and effective temperature were calculated with our stellar structure
code. This code was originally developed for the calculation of
finite-temperature mass-radius relations and equilibrium models for
pulsation studies (Koester \cite{koester78}; Dziembowski \& Koester
\cite{dziembowski}). This code is not fully evolutionary, but solves
the static stellar structure equations; The energy equation is
replaced by setting $l(r) \propto m(r)$. This is a sufficiently
accurate approximation, since in the outer layers it results in $ l
\approx L = const$ and in the interior the temperature becomes nearly
constant because of the high heat conduction of the degenerate
electrons.  The input physics -- equation of state and opacities --
have been updated and the outer boundary values are determined from
accurate DB atmospheric models. The parameterization of the mixing
length version is ML2, $\alpha = 0.6$ (see e.g. Fontaine et
al. \cite{fontaine81}, Tassoul et al. \cite{Tassoul}, or Koester et
al. \cite{koester94} for an explanation of nomenclature). The
mass-radius relation predicted by these models is in reasonable
agreement with more sophisticated calculations (e.g. Wood
\cite{wood}).

This code predicts the total mass in the convection zone as a function
of the mass and effective temperature of the star. By multiplying this
with the hydrogen abundance we obtain the total hydrogen mass within
the convection zone in solar masses.

\section{Results}

The properties and fit results of the DB stars without visible hydrogen in
their atmosphere are listed in Table \ref{table:1}. The data for the DBA stars
are shown in Table \ref{table:2}.

\subsection{Uncertainties}

\begin{figure}
\centering
\includegraphics[height=\columnwidth,angle=-90]{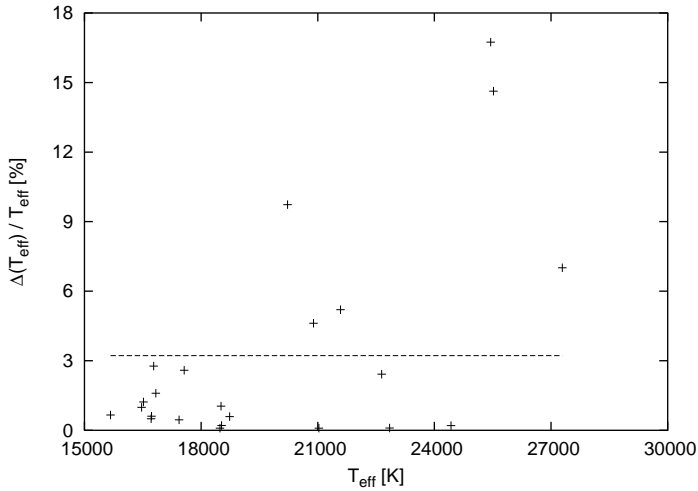}
\caption{The relative differences of the temperatures that we
derive for two spectra of the same object.}
\label{2_teff}
\end{figure}

\begin{figure}
\centering
\includegraphics[height=\columnwidth,angle=-90]{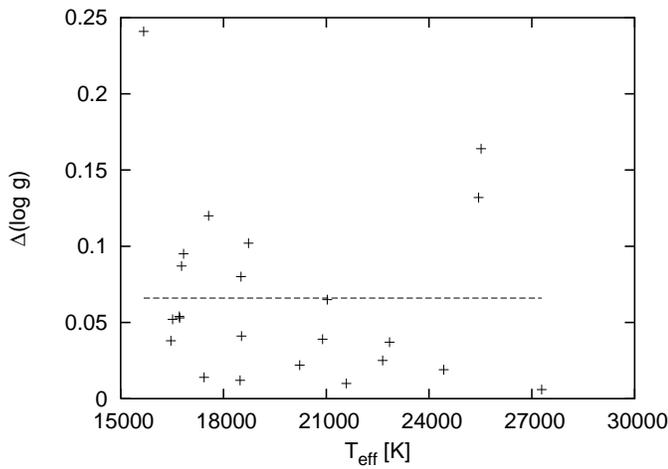}
\caption{The same as Fig. \ref{2_teff}, for $\log g$.}
\label{2_logg}
\end{figure}

The formal uncertainties from the $\chi^2$ fitting routine are mostly
low, and normally amount to only a few times 10\,K. However, they do
not include systematic errors, which might be introduced by the data
reduction, or by the lack of a flux calibration of the data. Thus the
true uncertainties of the fit results are generally underestimated by
the formal uncertainties. A better estimate can be obtained by
comparing the fit results of the two independent SPY spectra of each
object, an approach that was earlier used by Liebert et
al. \cite{liebert2005} in their study of DA white dwarfs. We have done
this for 24 of our objects for which both spectra are not too
different in quality and arrive at mean differences of $<\!\!\Delta
T_{\mathrm{eff}}\,/\,T_{\mathrm{eff}}\!\!>=3.22\,\%$ and $<\!\!\Delta
\log g\!\!>=0.066\,$. The large $<\!\!\Delta T_{\mathrm{eff}}\!\!>$ is
dominated by two outliers at high temperatures. If these are omitted,
the results are $<\!\!\Delta
T_{\mathrm{eff}}\,/\,T_{\mathrm{eff}}\!\!>=2.03\,\%$ and $<\!\!\Delta
\log g\!\!>=0.058\,$. These results are plotted in
Figs. \ref{2_teff} and \ref{2_logg}.

For six of the DBA objects we have conducted the full analysis for
both spectra and can thus also directly estimate the uncertainties of
H/He and $M_{\mathrm{H}}$; five of the six objects show very low
differences of abundance and hydrogen mass, $<\!\!\Delta
\log(\mathrm{H/He})\!\!>=0.02$ and $<\!\!\Delta
\log(M_{\mathrm{H}})\!\!>=0.025$, i.e., the total H masses differ by,
on average, 6\%. These estimates are of course based on a small number
of objects and thus only allow to judge the order of magnitude of the
uncertainties.

\begin{figure}
\centering
\includegraphics[height=\columnwidth,angle=-90]{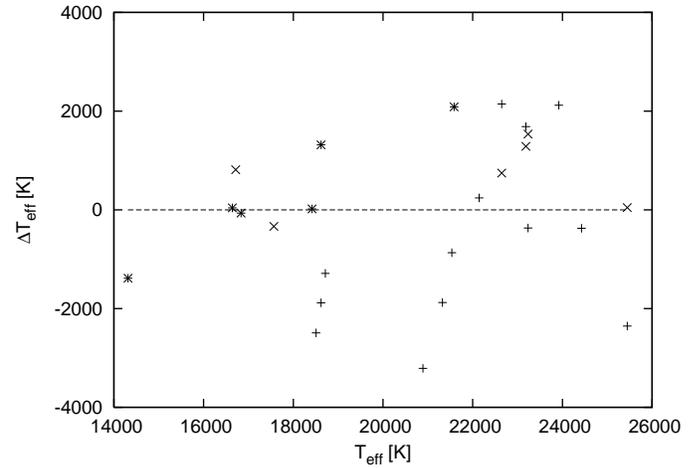}
\caption{The differences of the temperatures that we and other authors
  derive for common objects, plotted against our temperature. The
  '+'-symbols show the comparison to the results of Beauchamp et
  al. (\cite{beauchamp99}), the 'x'-symbols that to Castanheira et
  al. (\cite{castanheira}) and the asterisks plot the comparison to
  the values of Friedrich et al. (\cite{friedrich}).}
\label{comp_teff}
\end{figure}

\begin{figure}
\centering
\includegraphics[height=\columnwidth,angle=-90]{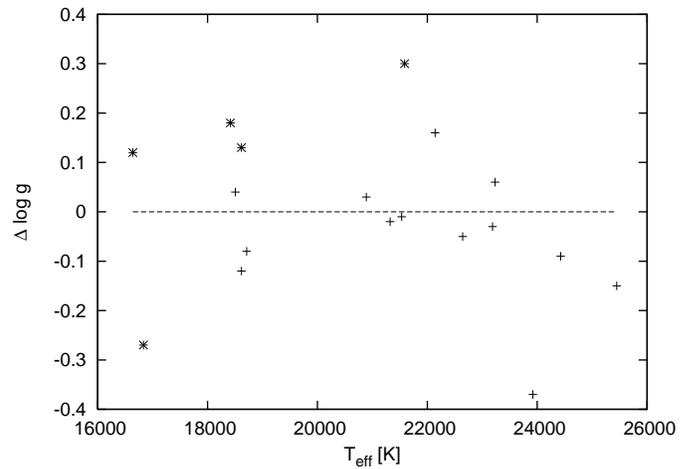}
\caption{The same as fig. \ref{comp_teff}, for $\log g$.}
\label{comp_logg}
\end{figure}

A comparison to the parameters that other authors derive for common
objects looks less satisfying; the differences in temperature between
our results and those of Beauchamp et al. (\cite{beauchamp99}),
Friedrich et al. (\cite{friedrich}), and Castanheira et
al. (\cite{castanheira}), shown in Fig. \ref{comp_teff}, are scattered
by more than $\pm10\%$ and the results of Beauchamp seem
systematically hotter by a few \%, or about 1500\,K. The surface
gravity differences, shown in Fig. \ref{comp_logg}, scatter by over
$\pm0.15$. It has to be noted that Castanheira et al.'s results are
derived from UV data, and that they use only pure helium
models. Therefore we only compare our results for DB without hydrogen
to their data. Beauchamp et al. {\em assumed} hydrogen abundances,
whereas we can use observed H lines. They assume H/He\,$=10^{-2}$ for
their DBA stars, and certain lower hydrogen abundances corresponding
to their detection limits for stars that they do not identify as
DBA. The systematic differences of our results to theirs may in part
arise because for {\em all} of our DBA stars that are in common with
Beauchamp's sample, we derive lower abundances than he assumed.

\begin{sidewaystable*}
\begin{minipage}[t][180mm]{\textwidth}
\caption{Properties and parameters of the DB stars without visible
  hydrogen. Two lines of data are given for the hot objects
  ($T_{\mathrm{eff}}>20\,000$\,K); the second line contains the result of a fit
  with H/He=$10^{-5}$ models.}  
\label{table:1}     
\centering
\renewcommand{\footnoterule}{} 
\centering                     
\vspace{1ex}
\begin{scriptsize}
\begin{tabular}{l l r l l r r r l}  
\hline\hline                 
Object & RA & DE & mag & Alias & $T_{\mathrm{eff}}$ [K]& $\log g$ & $M_*$ [$M_{\odot}$] & Comments \\ 
\hline                        
WD\,0119$-$004 & 01:21:48.27 & $-$00:10:54.4 & $V=16.0$ & G\,271$-$047A,
LP\,587$-$44, SDSS & 16\,542 & 8.0 & &\footnotemark[4]\\
WD\,0127$-$311 & 01:29:56.11 &$-$30:55:09.6 & $V=14.51$ & HE\,0127$-$3110,
GD\,1363 & 11\,002 & 8.0 & & only \ion{He}{i} 5876\,\AA\footnotemark[3]\\
MCT0149$-$2518 & 01:51:59.62 & $-$25:03:15.1 & $B=15.91$ & PHL\,1201 & 16\,834 & 7.940 & 0.557 &\\
WD\,0158$-$160 & 02:00:56.69 & $-$15:46:09.7 & $B=14.38$ & G\,272$-$B2B, MCT & 25\,518 & 7.875 & 0.542 &\\
 & & & & & 25\,259 & 7.891 & & \footnotemark[1]\\
WD\,0203$-$181 & 02:05:23.98 &$-$17:53:25.8 & $V=16.0$ & HE\,0203$-$1807,
G\,272$-$152 & 10\,757 & 8.0 & & only \ion{He}{i} 5876\,\AA\footnotemark[3]\\
WD\,0249$-$052 & 02:52:15.55 & $-$05:02:32.5 &  $B=16.1$ & HE\,0245$-$0514, KUV & 17\,908 & 7.992 & 0.589 & \\
HE\,0308$-$5635 & 03:09:47.84 & $-$56:23:49.6 & $B=14.02$ & WD\,0308$-$565,
BPM\,17088 & 22\,849 & 8.060 & 0.639 & \\
 & & & & & 22\,822 & 8.081 & & \footnotemark[1]\\
WD\,0349+015 & 03:51:51.65 & +01:39:47.6 & $B=16.3$ & KUV\,03493+0131 & 18\,741 & 7.889 & 0.535 & \\
HE\,0417$-$5357 & 04:19:10.04 & $-$53:50:45.7 & $B=15.12$ & BPM\,17731 & 18\,733
& 7.942 & 0.563 & \\
HE\,0420$-$4748 & 04:22:11.38 & $-$47:41:42.3 &  $B=14.7$ & & 27\,288 & 7.808 & 0.512 &\\
 & & & & & 27\,245 & 7.809 & &\footnotemark[1]\\
HE\,0423$-$1434 & 04:25:51.72 & $-$14:27:52.0 & $B=16.21$ & & 16\,904 & 7.794 &
0.481 &new object\\
HE\,0429$-$1651 & 04:32:13.81 & $-$16:45:08.5 & $B=15.82$ & & 15\,659 & 7.659 &
& DB+dM; new object\\
WD\,0615$-$591 & 06:16:14.48 & $-$59:12:28.1 & $V=14.09$ & L\,182$-$61, BPM\,18164 & 16\,714 & 8.017 & 0.600 &\\
WD\,0845$-$188 & 08:47:29.52 & $-$18:59:50.7 & $V=15.55$ & LP\,786$-$06, L\,0748$-$70 & 17\,566 & 7.969 & 0.575 &\\
WD\,0900+142 & 09:03:31.31 & +14:00:49.2 & $B=16.48$ & PG\,0900+142 & 15\,678 & 8.0 &  &\footnotemark[4]\\
WD\,1004$-$178 & 10:07:07.91 & $-$18:05:25.8 & $V=15.6$ & EC\,10047$-$1750 & 16\,357 & 7.700 & & DB+dM\\  
WD\,1046$-$017 & 10:48:32.65 & $-$02:01:11.3 & $V=15.81$ & GD\,124, GR\,387 & 15\,138 & 8.0 & &\footnotemark[4]\\
WD\,1144$-$084 & 11:46:54.07 & $-$08:45:48.5 & $B=15.95$ & PG\,1144$-$085 & 16\,468 & 8.0 & &\footnotemark[4]\\
WD\,1252$-$289 & 12:54:55.30 & $-$29:11:54.6 & $V=15.85$ & EC\,12522$-$2855 & 21\,029 & 7.97 & 0.595&\\
WD\,1326$-$037 & 13:29:16.37 & $-$03:58:51.8 &  $V=15.6$ & PG\,1326$-$037 & 22\,645 & 7.999 & 0.603& \footnotemark[2]\\
 & & & & & 22\,192 & 7.995 & & \footnotemark[1]\\
WD\,1336+123 & 13:39:13.63 & +12:08:29.6 &  $B=13.9$ & LP\,498$-$26 & 16\,779 & 8.0 & &\footnotemark[4]\\
WD\,1428$-$125 & 14:31:39.64 & $-$12:48:55.5 & $V=15.98$ & EC\,14289$-$1235 & 21\,586 & 8.167 & 0.701&\\ 
 & & & & & 21\,657 & 8.179 & &\footnotemark[1]\\
WD\,1445+152 & 14:48:14.37 & +15:04:49.6 & $B=15.55$ & PG\,1445+153 & 23\,234 &
8.012 & 0.611 &\\
 & & & & & 24\,063 & 7.968 & &\footnotemark[1]\\
WD\,1542+182 & 15:44:19.46 & +18:06:43.6 & $V=14.72$ & GD\,190, EG\,193 &
23\,186 & 7.966 & 0.585 &\\
 & & & & & 22\,902 & 7.965 & &\footnotemark[1]\\
WD\,1612$-$111 & 16:15:23.96 & $-$11:18:29.5 & $V=15.53$ & GD\,198, EG\,194 & 22\,143 & 8.073 & 0.645&\\ 
 & & & & & 22\,338 & 8.076 & &\footnotemark[1]\\
WD\,1654+160 & 16:56:57.56 & +15:56:26.6 & $B=16.15$ & PG\,1654+160, V\,824
Her & 25\,447 & 7.846 & 0.527 & V777\,Her variable\\ 
 & & & & & 25\,501 & 7.839 & &\footnotemark[1]\\
WD\,2129+000 & 21:32:16.39 &+00:15:13.3 & $V=14.73$ & LP\,638$-$4, PHL\,028
& 14\,414 & 8.0 & &\footnotemark[4]\\
WD\,2144$-$079 & 21:47:37.34 & $-$07:44:13.1 & $V=14.82$ & G\,026$-$031, LTT\,8702 & 16\,518 & 7.895 & 0.532 & DBZ\\ 
WD\,2234+064 & 22:36:42.06 & +06:40:16.4 & $y=16.26$ & PG\,2234+064, GR\,907 &
20\,890 & 8.035 & 0.620 &\\ 
 & & & & & 19\,864 & 8.028 & &\footnotemark[1]\\
WD\,2354+159 & 23:56:34.74 & +16:15:39.8 & $B=15.78$ & PG\,2354+159 & 24\,428 &
8.180 & 0.713 & DBZ \\
 & & & & & 22\,822 & 8.218 & &\footnotemark[1]\\
WD\,2354$-$305 & 23:56:37.07 & $-$30:16:25.4 & $B=16.26$ & & 18\,006 & 7.883 &
0.530 &\\
\hline
\end{tabular}
\end{scriptsize}
\footnotetext{$^{1}$\,Fit with H/He $=10^{-5}$ models to check the
influence of possible admixtures of invisibly small amounts of
hydrogen.}  
\footnotetext{ $^{2}$\,Beauchamp et al. (\cite{beauchamp99}) list
WD\,1326$-$037 as a DBA; however no H is visible in the SPY
spectra. According to Bergeron (2007, priv. comm.), the Beauchamp data
show no hydrogen features and thus this object appears as a DBA in
their paper due to a mis-classification.}
\footnotetext{$^{3}$\,Temperatures were
determined from the equivalent widths of \ion{He}{i} 5876\,\AA. The
surface gravity was assumed to be $\log g=8.0$.}
\footnotetext{$^{4}$\,Fitted with the alternate set of models with van
der Waals line broadening. The surface gravity was kept fixed at $\log
g=8.0$ for these fits.}
\end{minipage}
\end{sidewaystable*}

\begin{sidewaystable*}
\begin{minipage}[t][180mm]{\textwidth}
\caption{Properties and parameters of the DB stars with hydrogen features in
  their spectra. For some objects, two lines of data are given, for the
  analysis results of each of the two
  independent SPY spectra of the star.}     
\label{table:2}  
\centering
\renewcommand{\footnoterule}{} 
\centering                     
\vspace{1ex}
\begin{scriptsize}
\begin{tabular}{l l r r l r r r r r r l}  
\hline\hline                
Object & RA & DE & mag & Alias & $\log$ (H/He) & $T_{\mathrm{eff}}$
[K] & $\log g$ & $M_*$ [$M_{\odot}$] & Age [$10^8$ yr]& $M_{\mathrm{H}}$ [$M_{\odot}$] & Comments \\    
\hline                        
WD\,0000$-$170 & 00:03:31.64 & $-$16:43:58.6 & $V=14.67$ & G\,266$-$032,
L\,0793$-$018 & -5.61 & 13\,354 & 8.0 & 0.583 & 3.179 & 4.410$\times 10^{-12}$ &\footnotemark[6]\\
HE\,0025$-$0317 & 00:27:41.71 & $-$03:00:58.3 & $B=15.69$ & & -3.13 & 18\,570 &
8.293 & 0.777 & 2.017 & 4.765$\times 10^{-12}$\\
HE\,0110$-$5630 & 01:12:21.16 & $-$56:14:27.3 & $B=15.86$ & & -4.18 & 18\,483 &
8.120 & 0.666 & 1.350 & 5.535$\times 10^{-13}$ &DBAZ\\
 & & & & & -4.19 & 18\,465 & 8.132 & 0.673 & 1.393 & 5.585$\times 10^{-13}$ &\\
WD\,0125$-$236 & 01:27:44.57 & $-$23:24:49.3 & $y=15.42$ & G\,274$-$039, MCT &
-5.01 & 16\,757 & 8.065 & 0.629 & 1.780 & 1.287$\times 10^{-12}$ &\footnotemark[1]\\
HE\,0215$-$0225 & 02:17:32.67 & $-$02:11:15.0 & $B=16.13$ & (PB6822) & -5.67 &
16\,776 & 7.966 & 0.571 & 1.402 & 3.383$\times 10^{-13}$ &new object\\
WD\,0300$-$013 & 03:02:53.20 & $-$01:08:34.6 & $V=15.56$ & GD\,40, GR\,384 &
-6.02 & 15\,319 & 8.0 & 0.587 & 2.162 & 5.866$\times 10^{-13}$ &DBAZ\footnotemark[6] \\
HE\,0413$-$3306 & 04:15:20.63 & $-$32:59:10.0 &  $B=15.9$ & & -4.81 & 16\,640 &
7.915 & 0.543 & 1.284 & 3.088$\times 10^{-12}$ &\footnotemark[1]\\
HE\,0414$-$0434 & 04:16:52.65 & $-$04:27:23.6 & $V=15.7$ & & -5.20 & 13\,473 &
8.0 & 0.583 & 3.379 & 1.080$\times 10^{-11}$ & new object\footnotemark[6]\\
WD\,0503+147 & 05:06:15.83 & +14:48:31.1 &  $V=13.8$ & KUV\,05034+1445 & -4.91
& 16\,246 & 8.0 & 0.589 & 1.728 & 3.289$\times 10^{-12}$ &\footnotemark[1]$^,$\footnotemark[6]\\
WD\,0853+163 & 08:56:19.01 & +16:11:03.9 & $mc=15.83$ & PG\,0853+164, LB\,8827
& $\sim-3$ & 26\,128 & 8.396 & 0.853 & & &magnetic\footnotemark[2] \\
WD\,0921+091 & 09:23:55.27 & +08:57:17.5 & $V=16.19$ & PG\,0921+092, SDSS &
-4.81 & 18\,713 & 7.956 & 0.571 & 0.859 & 6.753$\times 10^{-14}$ &\\
WD\,0948+013 & 09:51:02.21 & +01:04:32.1 & $B=15.89$ & PG\,0948+013, SDSS &
-5.09 & 16\,939 & 7.967 & 0.572 & 1.354 & 1.061$\times 10^{-12}$ &\footnotemark[1]\\
WD\,1115+158 & 11:18:22.77 & +15:33:33.3 & $B=16.12$ & PG\,1115+158, DT\,Leo &
-2.44 & 23\,920 & 7.523 & 0.385 & 0.198 & 7.172$\times 10^{-17}$ &V777 Her variable\footnotemark[1]\\
WD\,1134+073 & 11:36:54.31 & +07:03:35.7 & $B=16.32$ & PG\,1134+073 & -4.70 &
18\,135 & 8.315 & 0.791 & 2.304 & 2.739$\times 10^{-13}$
&DBAZ\footnotemark[1]$^,$\footnotemark[4] \\
WD\,1149$-$133 & 11:51:50.61 & $-$13:37:15.1 & $V=16.29$ & EC\,11492$-$1320,
PG & -4.12 & 18\,618 & 8.196 & 0.714 & 1.570 & 4.807$\times 10^{-13}$ &\\
HE\,1207$-$2349 & 12:09:36.56 & $-$24:06:19.4 & $B=15.79$ & & -5.24 & 17\,435 &
7.935 & 0.556 & 1.107 & 3.578$\times 10^{-13}$ &new object\\
 & & & & & -5.26 & 17\,356 & 7.921 & 0.548 & 1.092 & 3.975$\times 10^{-13}$\\
EC\,12438$-$1346 & 12:46:30.42 & $-$14:02:40.8 & $V=16.39$ & & -4.63 & 16\,914 &
7.933 & 0.553 & 1.253 & 3.287$\times 10^{-12}$ &\footnotemark[1] \\
WD\,1311+129 & 13:13:51.35 & +12:40:09.4 & $mc=16.26$ & PG\,1311+129,
LP\,497$-$114, SDSS & -2.55 & 20\,224 & 7.890 & 0.539 & 0.519 & 4.878$\times
10^{-14}$ &\\
 & & & & & -1.95 & 22\,403 & 7.912 & 0.555 & 0.342 & 2.573$\times 10^{-15}$ &\\
WD\,1338$-$220 & 13:41:17.29 & $-$22:19:43.7 &  $B=16.2$ & HE\,1338$-$2204,
EC\,13385$-$2204 & -4.91 & 13\,694 & 8.0 & 0.584 & 3.210 & 1.925$\times
10^{-11}$ &\footnotemark[6]\\
HE\,1349$-$2305 & 13:52:44.26 & $-$23:20:06.5 & $B=16.36$ & & -4.67 & 18\,173 &
8.133 & 0.673 & 1.494 & 3.344$\times 10^{-13}$ & DBAZ\\
WD\,1352+004 & 13:55:32.38 & +00:11:24.1 & $B=15.72$ & PG\,1352+004, SDSS &
-5.08 & 14\,316 & 8.0 & 0.585 & 2.752 & 9.854$\times 10^{-12}$ & DBAZ\footnotemark[6]\\
WD\,1403$-$010 & 14:06:19.96 & $-$01:19:32.6 &  $V=15.9$ & G\,064$-$043,
GR\,272, SDSS & -5.21 & 15\,784 & 8.0 & 0.588 & 1.933 & 2.536$\times 10^{-12}$
& \footnotemark[6]  \\
HE\,1409$-$1821 & 14:11:48.67 & $-$18:35:05.6 & $B=15.6$ & & -5.14 & 18\,514 &
8.099 & 0.654 & 1.274 & 5.681$\times 10^{-14}$ & \footnotemark[1]\\
 & & & & & -5.32 & 18\,321 & 8.019 & 0.606 & 1.095 & 5.678$\times 10^{-14}$ &\\
WD\,1421$-$011 & 14:24:29.21 & $-$01:22:16.7 & $B=15.97$ & PG\,1421$-$011 &
-4.08 & 16\,788 & 7.997 & 0.589 & 1.503 & 1.220$\times 10^{-11}$ & DBAZ\footnotemark[1]$^,$\footnotemark[4]\\
WD\,1444$-$096 & 14:47:37.02 & $-$09:50:05.5 & $V=14.98$ & EC\,14449$-$0937,
PG1444$-$096 & -5.54 & 16\,724 & 8.044 & 0.617 & 1.710 & 4.066$\times 10^{-13}$
& \footnotemark[1]\\
 & & & & & -5.58 & 16\,622 & 7.989 & 0.584 & 1.540 & 4.712$\times 10^{-13}$
& \\
WD\,1456+103 & 14:58:32.74 & +10:08:17.6 & $B=15.89$ & PG\,1456+103, CW Boo &
-3.42 & 21\,533 & 7.904 & 0.549 & 0.404 & 2.443$\times 10^{-16}$ & V777 Her variable\\
WD\,1542$-$275 & 15:45:22.57 & $-$27:40:06.4 & $V=15.5$ & LP\,916$-$027 & -5.61 &
10\,826 & 8.0 & 0.577 & 6.789 & 7.422$\times 10^{-12}$ & only \ion{He}{i} 5876\,\AA~\& H$\alpha$\footnotemark[1]$^,$\footnotemark[3]\\
WD\,1557+192 & 15:59:21.08 & +19:04:09.0 &  $V=15.4$ & KUV\,15571+1913 & -4.88
& 18\,690 & 8.100 & 0.654 & 1.223 & 7.099$\times 10^{-14}$
& \footnotemark[1]\\
WD\,1709+230 & 17:11:55.68 & +23:01:01.5 &  $V=14.9$ & GD\,205, BPM\,92077 &
-4.36 & 18\,527 & 8.067 & 0.635 & 1.177 & 3.303$\times 10^{-13}$ & DBAZ\\
 & & & & & -4.41 & 18\,488 & 8.108 & 0.659 & 1.309 & 3.194$\times 10^{-13}$ & \\
WD\,1917$-$077 & 19:20:34.96 &$-$07:40:00.8 & $V=12.28$ & L\,923$-$021, LTT\,7658
& -5.16 & 10\,195 & 8.0 & 0.576 & 8.280 & 2.069$\times 10^{-11}$ & only \ion{He}{i} 5876\,\AA~\& H$\alpha$ \footnotemark[3]\\
WD\,2130$-$047 & 21:33:34.86 & $-$04:32:24.9 & $V=14.5$ & GD\,233,
L\,1002$-$062 & -5.34 & 17\,562 & 7.951 & 0.565 & 1.115 & 2.266$\times 10^{-13}$ & \footnotemark[1]\\
WD\,2142$-$169 & 21:45:27.60 & $-$16:43:08.0 & $B=14.6$ & HK\,22944$-$44 & $>-2$ & 39\,138
& 7.737 & 0.505 & & & DO\footnotemark[5]\\
WD\,2154$-$437 & 21:58:01.55 & $-$43:28:00.8 & $V=15.04$ & L\,427$-$060,
BPM\,44275 & -4.78 & 16\,734 & 8.018 & 0.601 & 1.601 & 2.472$\times 10^{-12}$ & \footnotemark[1]\\
WD\,2229+139 & 22:31:45.50 & +14:11:13.1 & $mc=15.99$ & PG\,2229+139, GR\,906
& -4.54 & 15\,423 & 8.0 & 0.587 & 2.107 & 1.607$\times 10^{-11}$ &\footnotemark[6]\\
HE\,2237$-$0509 & 22:39:40.75 & $-$04:54:17.2 & $V=14.0$ & & -4.80 & 12\,134 &
8.0 & 0.581 & 4.731 & 4.175$\times 10^{-11}$ & only \ion{He}{i} 5876\,\AA~\& H$\alpha$\footnotemark[3]\\
WD\,2253$-$062 & 22:55:47.43 & $-$06:00:51.4 & $y=15.06$ & GD\,243, EG\,230 &
-4.24 & 17\,380 & 7.981 & 0.581 & 1.256 & 3.779$\times 10^{-12}$ & \\
WD\,2316$-$173 & 23:19:35.39 &$-$17:05:28.8 & $V=14.04$ & G\,273$-$013,
L\,0791-040 & -5.27 & 10\,868 & 8.0 & 0.577 & 6.708 & 1.625$\times 10^{-11}$ & only \ion{He}{i} 5876\,\AA~\& H$\alpha$ \footnotemark[3]\\
HE\,2334$-$4127 & 23:37:38.74 & $-$41:10:32.7 & $B=15.61$ & (LTT\,9631) &
-5.34 & 18\,250 & 8.029 & 0.611 & 1.140 & 6.369$\times 10^{-14}$ & new object\\
\hline
\end{tabular}
\end{scriptsize}
\footnotetext{$^{1}$\,First reported as a DBA here; previously known as DB.}
\footnotetext{$^{2}$\,Since WD\,0853+163 is a magnetic DBA, the fit is
  uncertain because the Zeeman-splitted lines were fitted with a non-magnetic
  model.}
\footnotetext{$^{3}$\,Temperatures and hydrogen abundances where determined
  from the equivalent widths of \ion{He}{i} 5876\,\AA~and H$\alpha$, respectively. The
  surface gravity was assumed to be $\log g=8.0$.}
\footnotetext{$^{4}$\,WD\,1134+073 and WD\,1421$-$011 show \ion{Ca}{ii} lines but are
  not mentioned among the SPY DBZ/DBAZ objects of Koester et al. (\cite{koester2005b}).}
\footnotetext{$^{5}$\,WD\,2142$-$169 shows a \ion{He}{i}-dominated spectrum with Balmer lines and \ion{He}{ii} 4686\,\AA.}
\footnotetext{$^{6}$\,Fit using the alternate set of models with van der Waals
  line broadening. The surface gravity was kept fixed at $\log g=8.0$ for these
  fits.}
\end{minipage}
\end{sidewaystable*}

\subsection{Hot DB stars}

Low amounts of hydrogen are spectroscopically invisible in hot
atmospheres but can influence the shape of the helium lines
through a change of the stratification and line broadening and thus
affect the result of a model atmosphere fit. To test this effect we
have fitted the DB stars with $T_{\mathrm{eff}}>20\,000$\,K also with
models of atmospheres with an admixture of an invisibly small amount
of hydrogen, H/He\,$=10^{-5}$. These fit results are also listed in
Table \ref{table:1}. The surface gravities do not differ significantly
to those derived from hydrogen-free atmosphere models; the
temperatures are on average larger if hydrogen is mixed into the
atmosphere. The average temperature difference is $\overline{\Delta
T_{\mathrm{eff}}}=231$\,K, and thus on the order of the uncertainties
of the measurements. The unknown hydrogen abundance therefore adds
some additional uncertainty to the atmosphere parameters of hot DB
stars, but this effect is much less pronounced in our results
than in those of Beauchamp et al. (\cite{beauchamp99}) and Castanheira
et al. (\cite{castanheira}). This is not surprising because the higher
resolution of our spectra allows us to use an upper limit of the
abundance of spectroscopically invisible that is much lower than that
of the earlier investigations, who used upper limits between
H/He\,$=10^{-3.5}$ (Castanheira et al.) and H/He\,$=10^{-4}$
(Beauchamp et al.).

Our sample includes only two V777\,Her variables, and thus
our results do not allow any meaningful statement on the structure of
the V777\,Her instability strip.

\subsection{Peculiar objects}
\subsubsection{The Cool DO Star WD\,2142$-$169}
The spectrum of WD\,2142$-$169 is dominated by
\ion{He}{i} lines, but also shows a weak \ion{He}{ii} 4686\,\AA~line,
as well as H$\alpha$, H$\beta$, and \ion{Ca}{ii} lines. In a
classification determined by the relative strength of the spectral
features this star would be a DBO; however in accordance with
Eisenstein et al. (\cite{eisenstein_b}) we classify it as a DO rather
than a DB or DBO. The presence of \ion{He}{ii} indicates that the
temperature can not be much smaller than 40\,000\,K, but the weakness
of the line suggests that the temperature of WD\,2142$-$169 is lower
than that of most other DO stars, almost all of which have
$T_{\mathrm{eff}}>45\,000\,$K.

\begin{figure}
\centering
\includegraphics[height=\columnwidth,angle=-90]{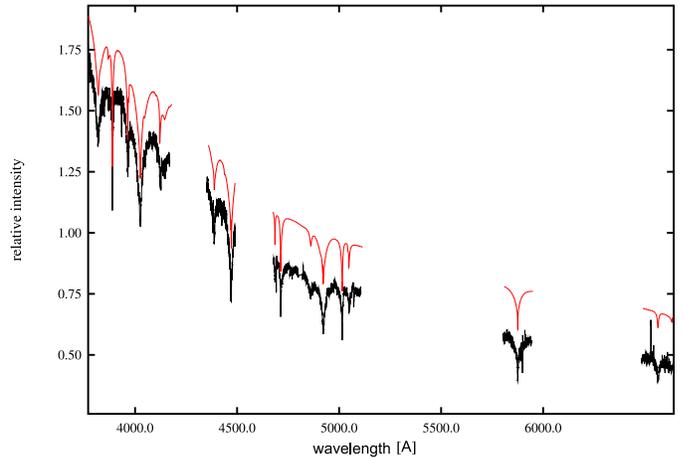}
\caption[The full spectrum of the cool DO star WD\,2142$-$169]{The
  fitted part of the spectrum of WD\,2142$-$169, shown as a black
  line, and the fit in red, shifted upwards by a constant amount for clarity.}
\label{WD2142fig}
\end{figure}

\begin{figure}
\centering
\includegraphics[height=\columnwidth,angle=-90]{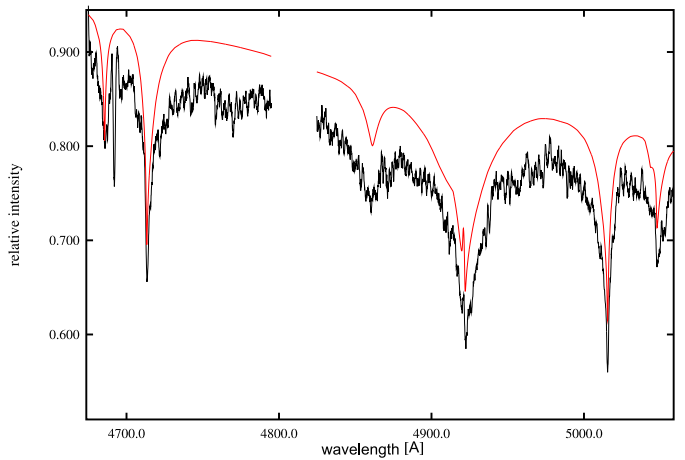}
\caption[The central part of the spectrum of the cool DO star
  WD\,2142$-$169]{The part of the spectrum of WD\,2142$-$169 that
  contains the  \ion{He}{ii} 4686\,\AA~and H$\beta$ lines. The spectrum is
  shown as a black line, and the fit in red, shifted upward. The feature
  at 4690\,\AA~is caused by a defect of the CCD that appears in many SPY
  spectra.}
\label{WD2142figseg}
\end{figure}

WD\,2142$-$169 is listed as a DB star of $T_{\mathrm{eff}}=15\,900\,$K
in Koester et al. (\cite{koester2005b}), and in fact such a fit can be
obtained at a moderate fit quality, and with a low hydrogen abundance
of H/He $=10^{-5}$, but of course such a fit does not reproduce the
\ion{He}{ii} line. The best fit is achieved with a much higher
hydrogen abundance, H/He $=10^{-2}$, and results in a temperature of
$T_{\mathrm{eff}}=39\,100\,$K and a gravity of $\log g=7.74$. This fit
is shown in Figs. \ref{WD2142fig} and \ref{WD2142figseg}. Note that
NLTE effects, which are important in hotter DOs, disappear below
40\,000\,K (Dreizler \& Werner \cite{DreizlerWerner}). The
\ion{He}{ii} line is slightly stronger in the fit than in the
spectrum, indicating that the atmosphere is perhaps slightly cooler
than the fit result. The Balmer lines are not well reproduced by this
fit; they are stronger in the observed spectrum than in the fitted
model, and thus the H abundance is probably higher than fitted, H/He
$>10^{-2}$. A fit with a model of a low hydrogen abundance usually
overestimates the temperature. Thus, both the \ion{He}{ii} line and
the hydrogen features point to a true temperature that is even
slightly lower than 39\,000\,K.

WD\,2142$-$169 is therefore one of the coolest DO stars that are known
and one of very few helium-rich objects that populate the DB gap. Only
one similar object is known so far, SDSS\,J074538.1+312205 (Eisenstein
et al. \cite{eisenstein_b}). Like WD\,2142$-$169, it shows weak
\ion{He}{ii} and Balmer lines in a spectrum that is dominated by
\ion{He}{i}. Eisenstein et al. derive a temperature of
39\,800$\pm$2000\,K from the SDSS photometry, however they do not
obtain a reliable fit of the objects' spectrum.

\subsubsection{WD\,0453$-$295}
This apparent DAB white dwarf was studied by Wesemael et
al. (\cite{wesemael94}) and found to be a DA+DB binary, since no
single star atmosphere model fit is able to reproduce the flux
distribution of this object.

\begin{figure}
\centering
\includegraphics[width=6.5cm,angle=-90]{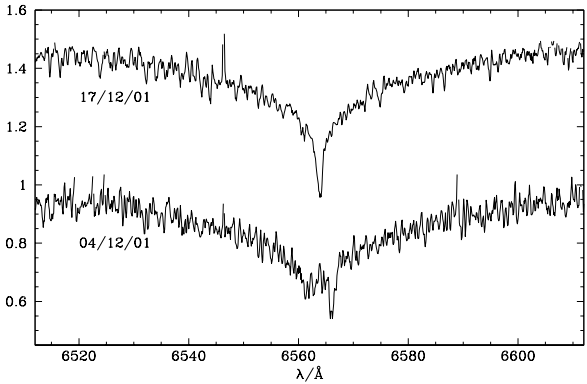}
\caption[The split H$\alpha$ line of WD\,0453$-$295]{The H$\alpha$ line in
  the two SPY spectra of WD\,0453$-$295. The lower spectrum shows two line
  cores. Taken from Napiwotzki et al. (\cite{Napi2005}).}
\label{WD0453fig}
\end{figure}

The SPY spectra of WD\,0453$-$295 (see Fig. \ref{WD0453fig}) show
variable line core shapes, due to the orbital motion of the
components. Napiwotzki et al. (\cite{Napi2005}) found that the core of
H$\alpha$ is split in one of the spectra. Thus the DB component is
indeed a DBA or perhaps even a DAB star, whose H abundance has to be
considerable, since the equivalent widths of H$\alpha$ of both the DA
and its companion have to be comparable.

\subsubsection{HE\,2149$-$0516}
The spectrum of HE\,2149$-$0516 shows strong Balmer lines and weaker
but strong \ion{He}{i} lines. It is a new DAB star. A pure helium (or
hydrogen) atmosphere fit does not give a result since the line
strengths are too strongly affected by the other constituent of the
atmosphere; only a pure hydrogen atmosphere fit that is limited to the
H$\alpha$ line results in a reasonable temperature estimate of
29\,000\,K to 30\,000\,K, but with a very low fitted gravity of $\log
g\approx6.9$. Alternatively, we find $T_{\mathrm{eff}}=32\,000\,$K if
the gravity is kept fixed at $\log g=8.0$. These fits of the line are
not good and the results are thus only very rough
estimates. Nevertheless, the resulting temperature is similar to those
that were derived for most other stars of this class, i.e., near the
low-temperature edge of the DB gap.

\subsection{DB and DBA mass distributions}

\begin{figure}
\centering
\includegraphics[height=\columnwidth,angle=-90]{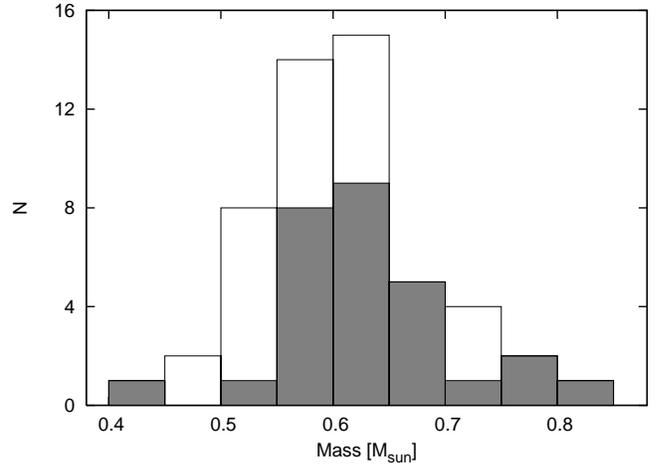}
\caption{The mass distribution of the DB stars. The filled part of each bin
  represents the DBA stars.}
\label{Figmasses}
\end{figure}

Fig. \ref{Figmasses} shows the mass distribution of the DB and DBA
stars. The magnetic DBA WD\,0853+163, the cool DO, and the objects for
which $\log g$ was kept fixed during the model atmosphere fit are not
included in this plot. The overall shape of the distribution is
similar to that which is found for the DA stars by, e.g., Liebert et
al. (\cite{liebert2005}), Giovannini et al. (\cite{Giovannini}), or
Madej et al. (\cite{madej}), as well as for the SPY DA stars (Voss et
al. \cite{voss}, in preparation). The mean mass of all stars is
0.596$\pm0.072\,M_{\odot}$, and thus very similar to the average mass
of the DA stars that has long been found to be about
0.55\,-\,0.6\,$M_{\odot}$. The similarity is especially striking for
the subsample of DBs that do not show any hydrogen in their spectra;
their mean mass is 0.584$\pm0.059\,M_{\odot}$. The mean mass of the
DBA subsample is only very slightly higher,
0.607$\pm0.082\,M_{\odot}$. Such a trend towards higher masses in DBA
stars was already found by Beauchamp et al. (\cite{beauchamp96}),
although much more pronounced: The mean masses for their sample of 13
DBA stars and 41 DB without spectroscopically visible hydrogen are
$<\!\!M\!\!>=0.642\,M_{\odot}$ for the DBA and
$<\!\!M\!\!>=0.567\,M_{\odot}$ for the DB. They speculated that the
presence of hydrogen could be favored by a larger stellar mass through
an increased hydrogen accretion from the interstellar medium, or
through a thinner convection zone in which hydrogen dilution would be
decreased.

However, the differences that we find between our DB and DBA
distributions are not significant. Furthermore, to test Beauchamp et
al.'s speculation, we tried to identify any dependency of the hydrogen
abundance, the hydrogen mass, or the hydrogen accretion rate on the
stellar mass. No such dependency is found, which adds confidence to
the assumption that DBAs and DBs are not intrinsically different from
each other.

Beauchamp et al. also found that the distribution of the DB masses is
narrower than that of the DA stars, only 5\% of their DB stars are
placed outside of a mass interval from 0.5\,$M_{\odot}$ to
0.65\,$M_{\odot}$. Here, a fraction of 25\% of the DBs are found
outside that interval, which is not too different from the 32\% of the
SPY DA stars that have masses outside this range.

Finally, the DB mean mass of this analysis is much lower than the
unexplainedly high value of 0.77\,$M_{\odot}$ that was found in paper
I. The difference is probably due to the fact that paper I used an
earlier, less sophisticatedly reduced version of the data. A major
improvement was a better correction of the response function of the
spectrograph in the blue spectra. The imperfect correction in the
previous version was likely responsibe for a significant fraction of
the systematic offset. The reason for the high paper I masses may be
not entirely clear, but it is reassuring that we now find a mass
distribution that agrees with that of the well studied 80\% of all
White Dwarfs, the DA. According to these results, there is no need to
assume any distinction between the structures of white dwarfs with
hydrogen- or helium-rich atmospheres, or, for that matter, between DB
and DBA stars.

\subsection{Hydrogen abundance and total hydrogen mass in the DBA stars}

\begin{figure}
\centering
\includegraphics[height=\columnwidth,angle=-90]{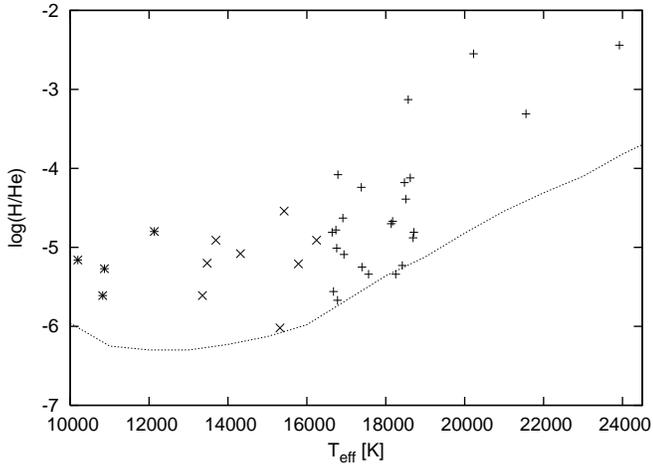}
\caption{The hydrogen abundances of the DBA stars as a function of the
  effective temperature. The tilted crosses show the objects for
  which the alternate model grids were used and for which the surface gravity
  was kept fixed during the fit. Asterisks show the stars for which
  temperatures and abundances were derived from the equivalent line widths of
  \ion{He}{i} 5876\,\AA~and H$\alpha$, respectively. The dotted
  line corresponds to the H$\alpha$ equivalent width detection limit
  of 300\,m\AA.}
\label{H/He}
\end{figure}

Fig. \ref{H/He} shows that the observed hydrogen abundances are lower
for cooler DBA stars than for the hottest objects. This decrease of
the average abundance is not unexpected since the helium convection
zone deepens with decreasing temperature, leading to increased
dilution of any hydrogen mass originally present.

The lowest observed H$\alpha$ equivalent widths in our DBA sample are
close to 300\,m\AA, and therefore we assume this value as the
detection limit for hydrogen in our data. The detection limit in terms
of abundance vs. temperature is shown by the dotted line in the
figure. The observed abundances reach near this limit at temperatures
around 18\,000\,K but are higher for the hottest and coolest
objects. This reflects the average quality of our spectra which is
somewhat lower for the hot and cool objects, and better at
intermediate temperatures.

\begin{figure}
\centering
\includegraphics[height=\columnwidth,angle=-90]{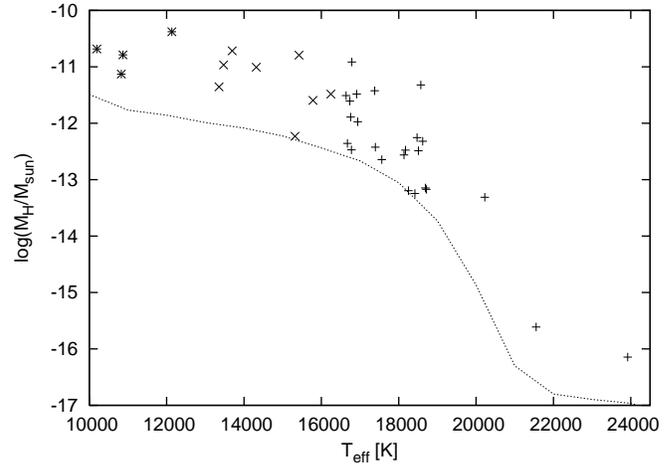}
\caption{The hydrogen masses plotted against the
  effective temperature; the symbols and the dotted line are the same as in
  fig. \ref{H/He}.}
\label{logMH_T}
\end{figure}

\begin{figure}
\centering
\includegraphics[height=\columnwidth,angle=-90]{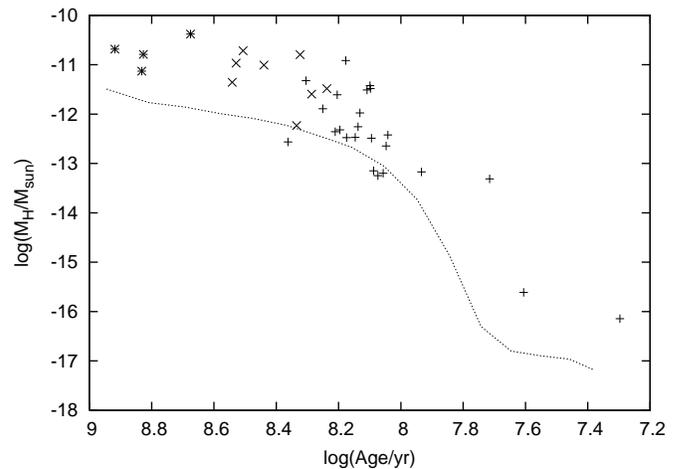}
\caption{The hydrogen masses plotted against log(Age);
  the symbols are again the same as in fig. \ref{H/He}. The distribution
  appears more scattered along the age axis than it is along the temperature
  axis in fig. \ref{logMH_T}. This is due to the effect of stellar mass on the
  cooling rate. The dotted line depicts the hydrogen detection limit for $\log
  g=8.0$. This 300\,m\AA-limit would be placed at higher ages for stars with
  $\log g>8.0$.}
\label{logMH_logAge}
\end{figure}

A more instructive quantity is the hydrogen mass itself, which is
shown in Fig.~\ref{logMH_T} as a function of temperature and in
Fig.~\ref{logMH_logAge} as a function of white dwarf age. Since we
assumed a constant mass below approx. 16\,000\,K, the qualitative
similarity between the two curves is not too surprising. Both of them
clearly show that the deduced H masses increase with the age of the
WD. Before interpreting this in a too simplistic way one should note,
however, the observational detection limit, which prevents any
observation of H masses below $10^{-12}\,M_{\odot}$ at the cool
end. On the other hand, the absence of larger H masses at the hot end
is explained easily: A WD with H mass of $10^{-12}\,M_{\odot}$ at
22\,000\,K would have a H/He number ratio $> 1$, and thus look more
like a DAB than a DBA.

\section{Discussion and conclusions}

The detection of low hydrogen abundances in such a large number of
spectra significantly increases the number of DBA stars among the
white dwarfs with helium-rich atmospheres. With a fraction of 55\%,
the majority of DBs in our sample are DBAs. The increase is not
unexpected because it seemed always plausible that decreasing the
detection limit would discover even weaker hydrogen lines.  There are
6 new objects in our sample which have not been reported before in the
literature, and 4 of these are DBAs. In addition to that, 14 stars of
our sample cataloged as DBs turn out to be DBAs. One of them is the
V777\,Her variable WD\,1115+158.

How do these observations fit with our understanding of DB evolution
and the DB gap? The evolution of a $0.6\,M_{\odot}$ DA in the DB gap
and its transformation into a DB by convective mixing has been studied
in great detail by MacDonald \& Vennes
(\cite{macdonaldvennes}). Depending on the convection description
used, the transition can occur between 25\,000 and 32\,000\,K, if the
hydrogen mass is $10^{-15}\,M_{\odot}$. For $10^{-14}\,M_{\odot}$ of H
mass these values go down to 11\,800 - 18\,000\,K. Since we observe
DBA white dwarfs up to 24\,000\,K in our sample, we assume $\approx
10^{-15}\,M_{\odot}$ as a reasonable starting value for those DA
within the DB gap, which will turn into hot helium-rich stars around
30\,000\,K. Because of the shallow convection zone this leads to the
high observed values of $\log$ H/He = -3 to -2; the apparently pure He
DBs in this temperature range may have just slightly lower H
masses. Because of the steep increase of the convective mass such
small amounts of hydrogen would become unobservable after a short
evolution to temperatures below 20\,000\,K.  We still find hydrogen in
these cooler stars, but now indicating much larger total H masses of
$\ge 10^{-13}\,M_{\odot}$. We can envisage two scenarios for the
origin of this hydrogen: Mixing of a thicker original H layer,
i.e. transformation of a DA into a DB at much lower temperatures than
the canonical 30\,000\,K, or interstellar accretion.

\begin{figure}
\centering
\includegraphics[height=\columnwidth,angle=-90]{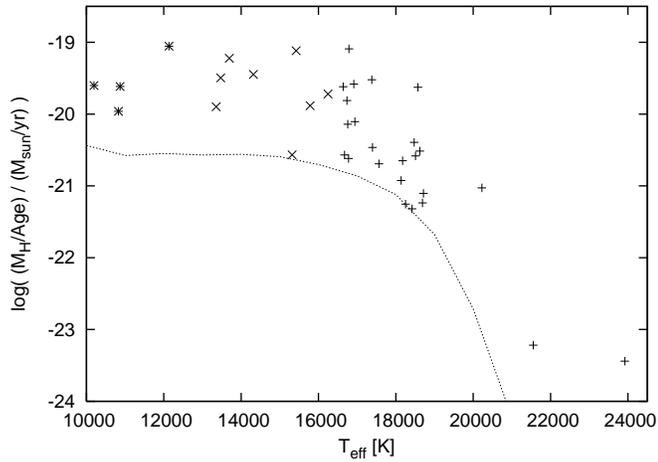}
\caption{The hydrogen accretion rate, shown on a logarithmic scale against the
effective temperature. The symbols are the same as in the previous figures.}
\label{logAccr_T}
\end{figure}

\begin{figure}
\centering
\includegraphics[height=\columnwidth,angle=-90]{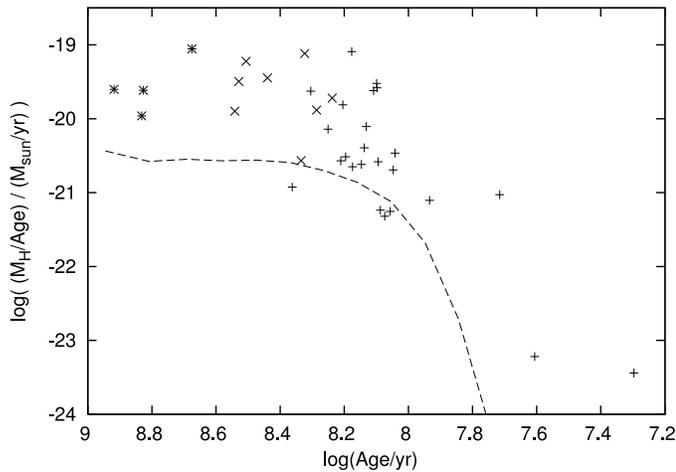}
\caption{The hydrogen accretion rate, shown on a logarithmic scale against the
age of the star. The symbols are the same as in the previous figures.}
\label{logAccr_Age}
\end{figure}

In view of the MacDonald \& Vennes (\cite{macdonaldvennes})
calculations we can discard the first scenario. None of their
different convection models allows mixing of $10^{-13}\,M_{\odot}$ at
a temperature higher than 12\,500\,K, whereas such masses are observed
already at 18\,000\,K. Assuming a constant accretion rate over the
lifetime of the DB, the accretion rates necessary to explain the
observations are extremely low: Between $10^{-19}$ and
$10^{-21}\,M_{\odot}$ per year (see Figs.\ref{logAccr_T} and
\ref{logAccr_Age}). In their important study of the
diffusion/accretion scenario for DZ white dwarfs, Dupuis et
al. (\cite{dupuis1993}) estimate a rate of $2.7 \times
10^{-22}\,M_{\odot}$ for a typical WD in the thin, warm phase of the
interstellar medium. This is based on the Eddington rate, using just
the gravitational cross section of a star in classical mechanics. It
is very hard to imagine a lower rate, unless some unknown mechanism
completely prevents accretion, but in any case such or even lower
rates would not lead to observable hydrogen. On the other hand, a
significantly higher rate -- continuous or episodic -- would lead to
H/He ratios of 0.01 to 0.1 at the low temperature end, as observed in
HS\,0146+1847 (Koester et al. \cite{koester2005a}) and GD362
(Zuckerman et al. \cite{zuckerman}). These stars appear at first sight
to be normal DAs, and only careful observation and analysis reveals
their unusual composition. It is currently completely unknown how many
more of these objects might be hidden among the ``normal''
DAs. Overall, our results thus strongly support the idea that small
but varying amounts of hydrogen are accreted over their lifetimes by
all DB white dwarfs.

The recent study by Dufour et al. (\cite{dufour2007}) of DZ white
dwarfs provides a very interesting continuation of our results towards
lower temperatures. In the overlap region between 10\,000 and
12\,000\,K, our results for the H/He ratios, total H masses, and
necessary accretion rates are in very good agreement. With some
precaution because of the selection effects, we can extend their
findings of the decreasing H accretion rate with effective temperature
much beyond their limit of 12\,000\,K up to 24\,000\,K. This may have
some relevance for one of the remaining mysteries, which is strongly
re-emphasized by Dufour et al. (\cite{dufour2007}): In those cases,
where metal traces are detected in addition to H (the DBZ, or DBZA),
the accretion rate for H compared to e.g. Ca is orders of magnitude
below that expected from solar abundances in the accreted matter. If
there is some kind of screening mechanism involved, which prevents or
hinders the accretion of hydrogen, it seems to increase its efficiency
in a rather smooth way towards high temperatures. This would argue
against the propeller mechanism (Wesemael \& Truran
\cite{wesemaeltruran}), which is most often invoked as a possible
explanation. The only other possibility we are aware of is the
existence of a weak stellar wind in hot DBs, which dies down with
decreasing temperature. In their wind model to explain the carbon
abundance in DBs, Fontaine et al. (\cite{Fontaine05}) {\em assumed} a
linear relationship between wind strength and age, and that it dies at
20\,000\,K.  Whether such a very weak wind (comparable in absolute
value to our derived accretion rates) exists down to very much lower
temperatures, and whether it can explain the difference in the Ca/H
accretion rates remains to be studied.

\begin{acknowledgements}
B.V. and D.K. acknowledge support by the Deutsche
Forschungsgemeinschaft (DFG) under project numbers KO738/21-1,
KO738/22-1, and KO738/23-1.
\end{acknowledgements}

\begin{appendix}
\section{Wavelength region around H$\alpha$ in the DBAs}

\begin{figure*}[hp]
\centering \includegraphics[height=17cm,angle=-90]{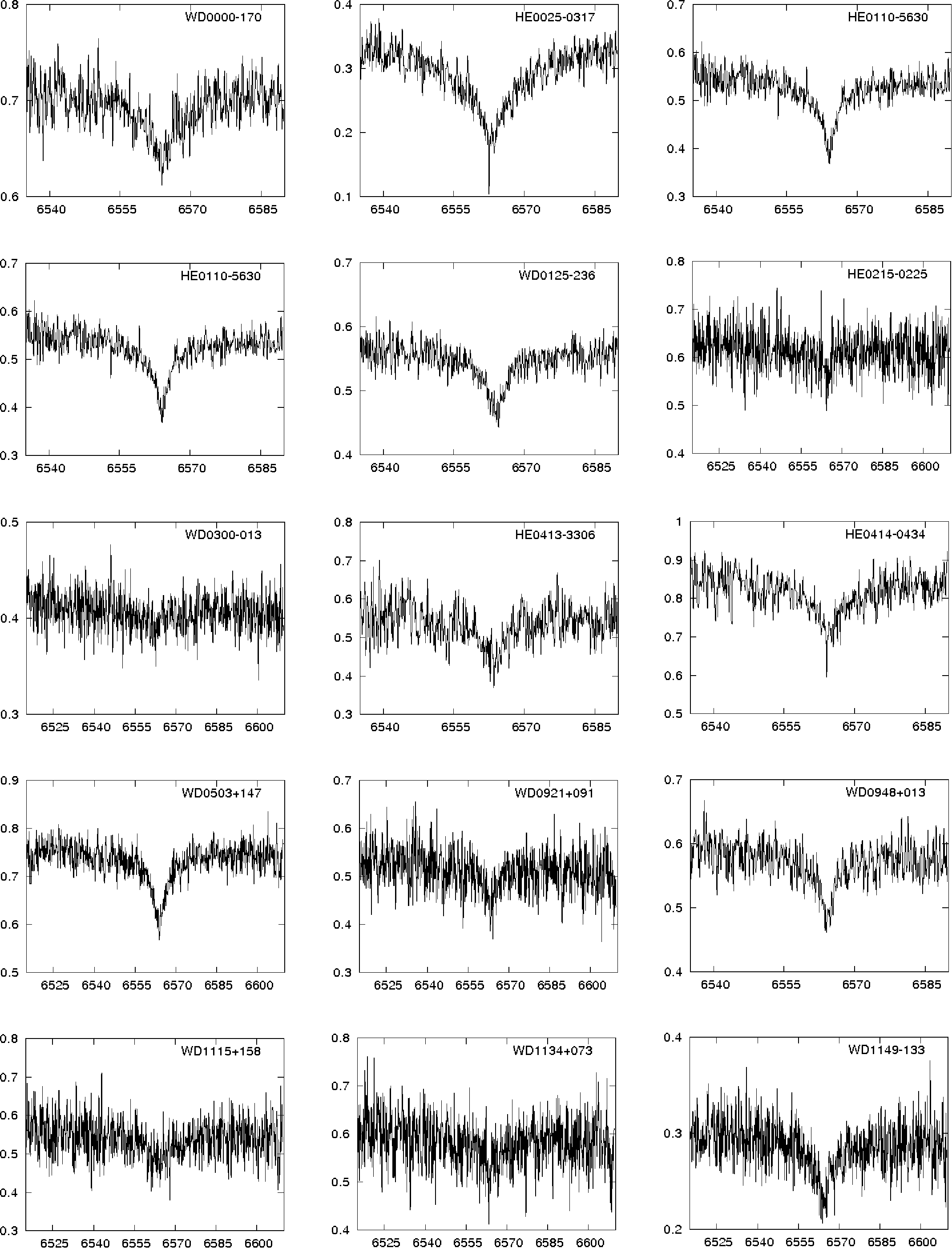}
\caption{{\bf The spectral regions around H$\alpha$ for all DBA stars,
continued in Figs. \ref{AllDBA2} and \ref{AllDBA3}. The intensity is
given in arbitrary units; the wavelengths are in \AA. Note that the
wavelength range is not the same in all plots.}}
\label{AllDBA}
\end{figure*}

\begin{figure*}[h!!p]
\centering \includegraphics[height=17cm,angle=-90]{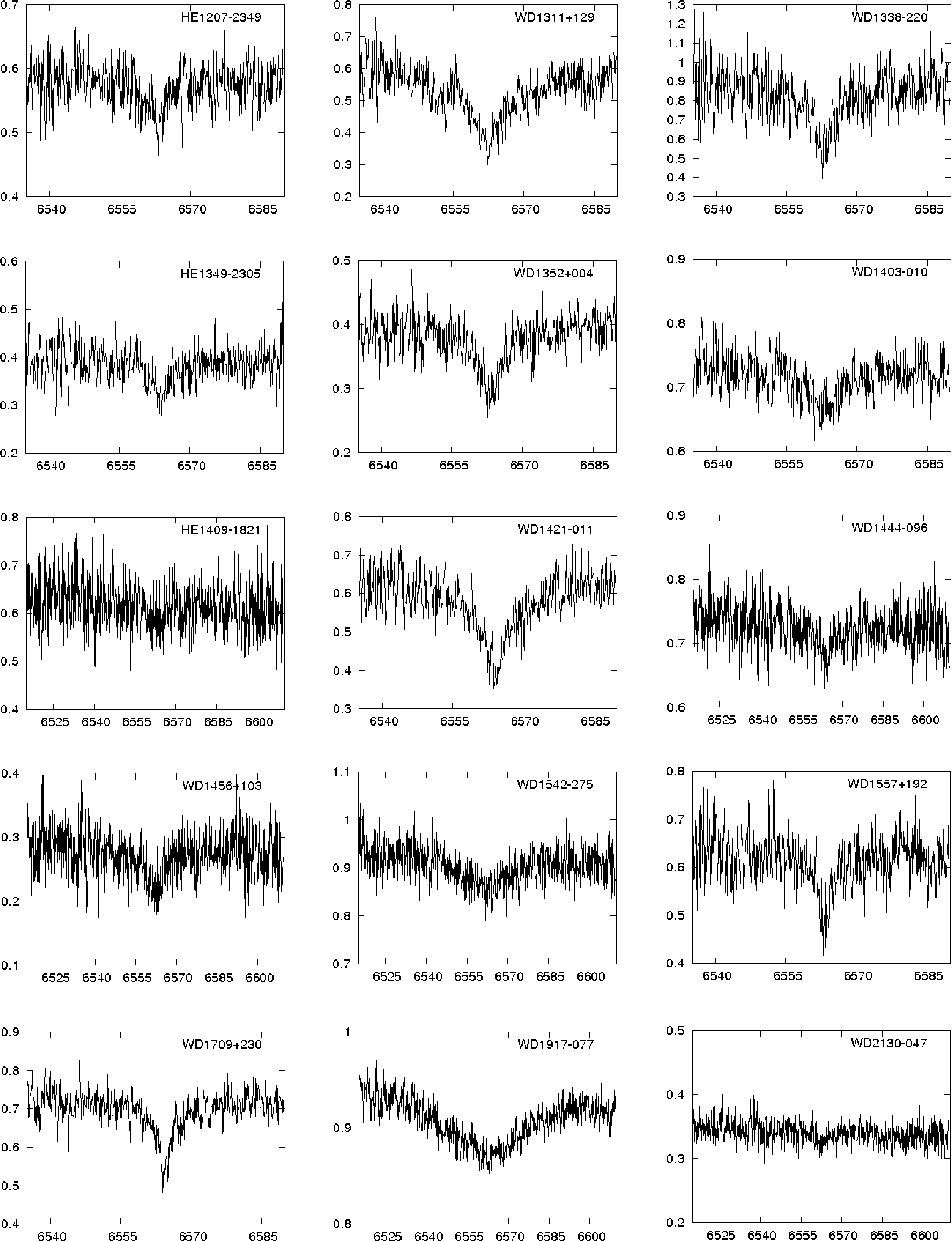}
\caption{{\bf The spectral regions around H$\alpha$ for all DBA stars,
continued from Fig. \ref{AllDBA}.}}
\label{AllDBA2}
\end{figure*}

\begin{figure*}[h!!t]
\centering
\includegraphics[height=17cm,angle=-90]{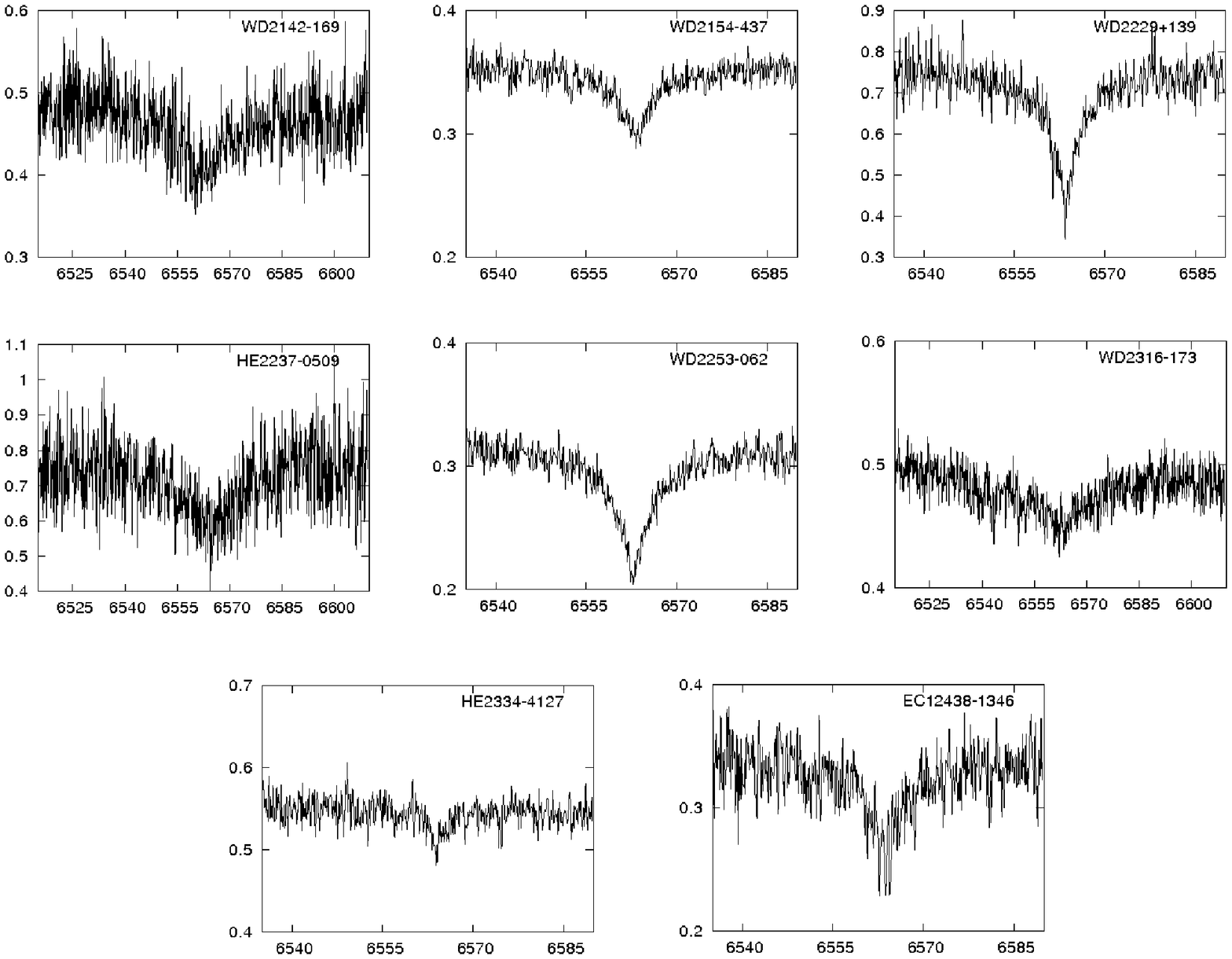}
\caption{{\bf The spectral regions around H$\alpha$ for all DBA stars,
continued from Figs. \ref{AllDBA} and \ref{AllDBA2}.}}
\label{AllDBA3}
\end{figure*}

\end{appendix}

\end{document}